\documentclass[aps,prl,twocolumn,superscriptaddress]{revtex4-1}

\usepackage[english]{babel}
\usepackage[utf8]{inputenc}
\usepackage{graphicx}
\usepackage{latexsym,amsmath,verbatim}
\usepackage{amssymb} 
\usepackage{amsfonts}
\usepackage{colortbl}
\usepackage{array}
\usepackage{stackrel}
\usepackage{bm}
\usepackage{nicefrac}
\usepackage[utf8]{inputenc}
\usepackage{rotating}
\usepackage{float}
\usepackage[final]{pdfpages}

\makeatletter
\AtBeginDocument{\let\LS@rot\@undefined}
\makeatother

\newcommand{\red}[1]{{ #1 }}
\newcommand{\new}[1]{{#1}}

\begin{document}

\title{Emergent spatial patterns of coexistence in species-rich plant communities}

\author{Pablo Villegas*}
\author{Tommaso Gili}
\affiliation{IMT Institute for Advanced Studies, Piazza San Ponziano 6, 55100 Lucca, Italy.}
\author{Guido Caldarelli}
\affiliation{Department of Molecular Sciences and Nanosystems, Ca’ Foscari University of Venice, 30172 Venice, Italy}
\affiliation{European Centre for Living Technology, 30124 Venice, Italy}
\affiliation{Institute for Complex Systems, Consiglio Nazionale delle Ricerche, UoS Sapienza, 00185 Rome, Italy}
\affiliation{London Institute for Mathematical Sciences, W1K2XF London, United Kingdom}

\vspace{0.5cm}

\vspace{0.5cm}

\renewcommand\refname{} 

\begin{abstract}
Statistical Physics has proved essential to analyze multi-agent environments. Motivated by the empirical observation of various non-equilibrium features in Barro Colorado and other ecological systems, we analyze a plant-species abundance model \new{of neutral competition},  presenting analytical evidence of scale-invariant plant clusters and non-trivial emergent modular correlations. Such first theoretical confirmation of a scale-invariant region, based on percolation processes, reproduces the key features in \new{natural rainforest} ecosystems and can confer the most stable equilibrium for ecosystems with vast biodiversity.
\end{abstract}


\maketitle


\section*{Introduction}

The stable coexistence of multiple ecological species represents a primary problem in theoretical ecology. Mesoscopic descriptions and agent-based modeling --aiming to elucidate the mechanisms upscaling the response of individuals (trees) to the ecosystem level-- have proven that constitutes a multi-scale problem far from being resolved \cite{Villa2015, Levin1992, Wiegand2021}. In this respect, the first theoretical problem arising in the study of extensive forests is defining the mechanisms determining the response at the ecosystem level from the individual one (trees) \cite{Levin1992}. \red{Usual approaches relying on ‘mean-field’ approximations --e.g., considering multiple interdependent Langevin's equations to determine individual average density for a single species \cite{Borile2012}-- indicate that stable coexistence is difficult to reach in large communities. However, they represent a crucial step demonstrating that detailed balance is not fulfilled \cite{VanKampen} (even though some analytical treatments are derived under such assumption \cite{Od2010}), making it possible to spontaneously broke the neutral symmetry \cite{Borile2012}.}


\new{We deal with complex aggregates in rainforests where a single species' presence is generally not expected.} Ecosystems with a high diversity level have been addressed so far within the Neutral Theory (NT) framework of biodiversity \cite{Hubbell2001, Azaele2016}. Few parameters are enough to describe \new{species abundance distribution (SAD)}, species-area relationships (SAR), and the main biodiversity indices in tropical forests \cite{Azaele2016, Rosindell2011}.
However, NT does not consider essential features like competition for resources nor how key ecological patterns (spatial tree patterns, SAR, and \new{SAD}) are intimately intertwined and scale-dependent, thus neglecting essential space-dependent aspects \cite{Azaele2016, Gravel2006}.

Adding space-dependent aspects, e.g., through Spatially Explicit Neutral Models (SENM), is equally problematic \cite{Rosindell2011}. \new{SENMs} have made progress in studying natural ecosystems \cite{Pigolotti2018, Pigolotti2009, Rosindell2007} by explaining patterns such as beta-diversity \cite{Chave2002} and species–area relationships \cite{Rosindell2007,Pigolotti2009}. 
Nonetheless, they also deal with many significant theoretical challenges concerning their non-spatial formulations \cite{Azaele2016,Grilli2012,Vergnon2012}.

The interdependence between the emergent spatial patterns --at the mesoscale-- and species coexistence has been often overlooked \cite{Wiegand2021}, even if their joint action with seed dispersal may be critical to ensure a rich biological diversity \cite{Detto2016, Bolker1999, Uriarte2010}. In particular, only recent works have emphasized the difficulty of \new{integrating} spatial patterns into coexistence theories of species-rich communities, showing \new{how they play an essential} role in species coexistence of high diversity plant communities \cite{Wiegand2021}. 

Some works have recently highlighted the ability of SENM to produce single-species patterns characterized by spatial density correlations and fluctuations in qualitative agreement with field data in the 50 ha tropical forest plot on Barro Colorado Island (BCI) \cite{Villegas2020, May2015, Grilli2012}. In particular, \new{it has been proved} the prevalence of tree clumping at small scales together with anomalous spatial density fluctuations \cite{Condit2000,Villegas2020,Keil2010}. Nevertheless, the role of dispersion and immigration in maintaining biodiversity richness and their effects in the emergent spatial point patterns remains an open question. Indeed, it still needs to be clarified whether correlations between species operate when competing for space. This is particularly important to describe patterns with characteristic scales (e.g., stripes or Namibian fairy circles \cite{Klausmeier1999,Tarnita2017}) or conversely lacking scale (e.g., the low-canopy gaps in BCI \cite{Sole1995} or tree canopy clusters across different Kalahari landscapes \cite{Scanlon2007}). Bistability associated with discontinuous transitions has been suggested to play a critical role in regular patterns in arid ecosystems \cite{Kefi2010} and percolation phenomena to justify the emergence of very broadly distributed vegetation patterns \cite{Martin2020}. Therefore, the emergence of scale-free clusters in multi-agent competitive environments, in concomitance with complex spatial patterns, remains a crucial question to be answered.

We propose here such a dynamic regime. For that purpose, we show the existence of a rich phase diagram in the SENM, where bistable effects, clustered patterns, and scale-invariant regions are present. In particular, scale invariance allows \new{many} species to live together, maintaining high specific spatial correlations, while the percolating nature of the spatial distribution of the intertwined species reveals a non-trivial emergent modular structure between them. This region constitutes the optimal regime for natural systems of exceedingly high biodiversity to harbor complex behavior, featuring optimal trade-offs between species richness, displaying a non-trivial emerging modular structure, and long-range spatial correlations at the single-species level.

\section*{Spatially Explicit Neutral Model (SENM)}

The investigation presented consists of a careful analysis of the phase diagram in a plant species abundance model's parameter space. We modeled this data through the Spatially Explicit Neutral Model (SENM) \cite{Hubbell2001} or multispecies voter model \cite{Durrett1996, Pigolotti2018} with a particular rule for the species seed dispersal. 
The SENM dynamics --on a square lattice of size $N=L^2$-- is defined as follows. A random tree gets replaced via: (i) speciation processes or immigration, i.e., introducing a new species with probability $\nu$, or (ii) dispersal effects, i.e., selecting another random tree from its neighborhood, with probability $1-\nu$. For the sake of simplicity, the neighborhood is defined as a square dispersal kernel of size $K$ centered on the gap (see a detailed description of the model in Supplementary Information (SI1)). The assumptions underlying the model are based on three main aspects: dispersal limitation, demographic fluctuations, and competition between species,  which are {\em a priori} equivalent at the individual level: i.e., they compete for space with identical birth/death rates and dispersal mechanisms \cite{Hubbell2001}.

\begin{figure}[hbtp]
 \begin{center}
    \includegraphics[width=0.7\columnwidth]{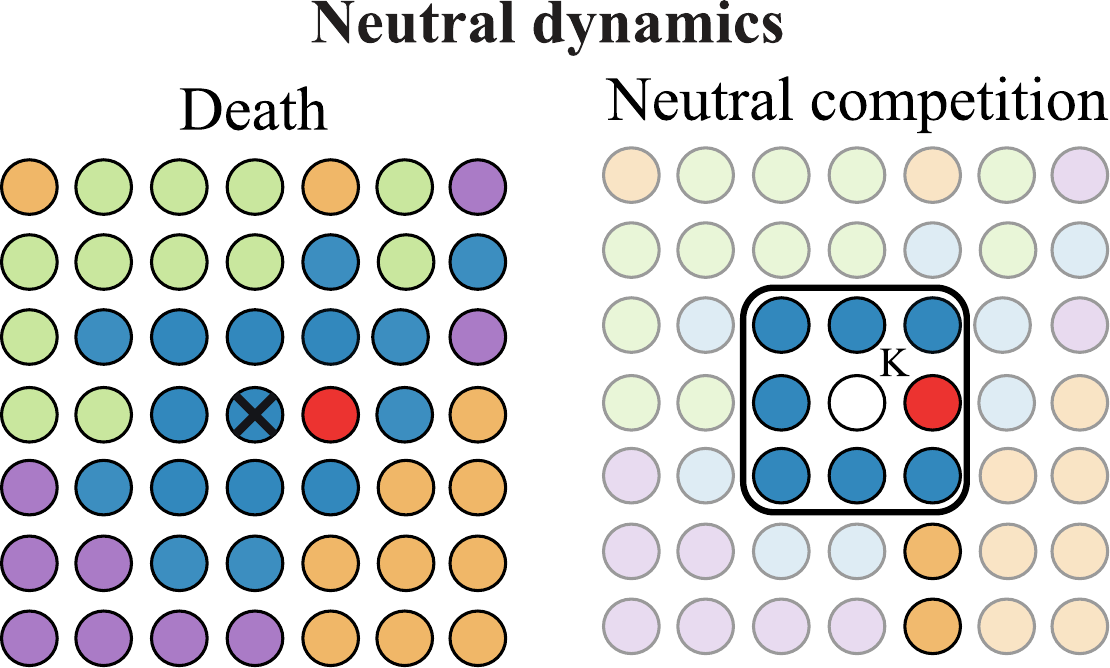}
    \caption{ \new{Sketch of the neutral dynamics. A place is selected, and the tree dies, being replaced by: (a) seed dispersal from a kernel K, or (b) immigration/speciation, with a small probability $\nu$.}}
     \label{NSketch}
\end{center}
\end{figure}


\paragraph{\textbf{Dual representation of the SENM}} 
\new{The simulations of the SENM have been made exploiting its duality with a system of coalescing random walkers \cite{Bramson1996, Durrett1996, Holley1975}. This widely used and powerful method has the great advantage of allowing high-speed simulations of independent realizations virtually free from boundary effects as, e.g., those of periodic boundary conditions (typically employed in the model's forward dynamics). Of course, this procedure can only be used if one is interested in the static, long-term properties of the model, and consequently, this duality enables the study of the ergodic properties of the infinite system \cite{Holley1975}. In particular, the generation of many sample patterns at the (non-equilibrium) stationary state allows exploration of the statistical ensemble.

The dual dynamics of the SENM works as follows (see also \cite{Pigolotti2009, Pigolotti2018} for an accurate description of the process): (i)  A random walker is placed on each lattice site, and the process proceeds backward in time to reconstruct the ancestry of the species. (ii) With probability $1-\nu$, a randomly chosen walker is moved --at each discrete (backward) time step-- to a different site selected from a squared dispersal kernel of length $K$ (pay particular attention to the possibility of choosing a site outside the sampled domain, since we only observe a portion of the infinite lattice).  If the landing site is occupied, the two walkers coalesce, remove one of them, and trace the coalescing partner. (iii) With probability $\nu$, the random walker is killed. This corresponds --in the forward dynamics-- to a speciation/immigration event associated with a new species' growth. (iv) The simulation finishes when only one walker remains. 

The complete history of coalescing events, i.e., the tree of coalescences, allows us to trace back the entire genealogical tree of a species up to the speciation event that originated it. In other words: on a dual simulation, once all walkers coalesced or were annihilated, species are assigned to the starting site of each walker, obtaining a possible configuration (i.e., a 'snapshot' from the ensemble of possible states) of the SENM \cite{Pigolotti2009, Pigolotti2018}.}

\begin{figure*}[hbtp]
 \begin{center}
    \includegraphics[width=1.8\columnwidth]{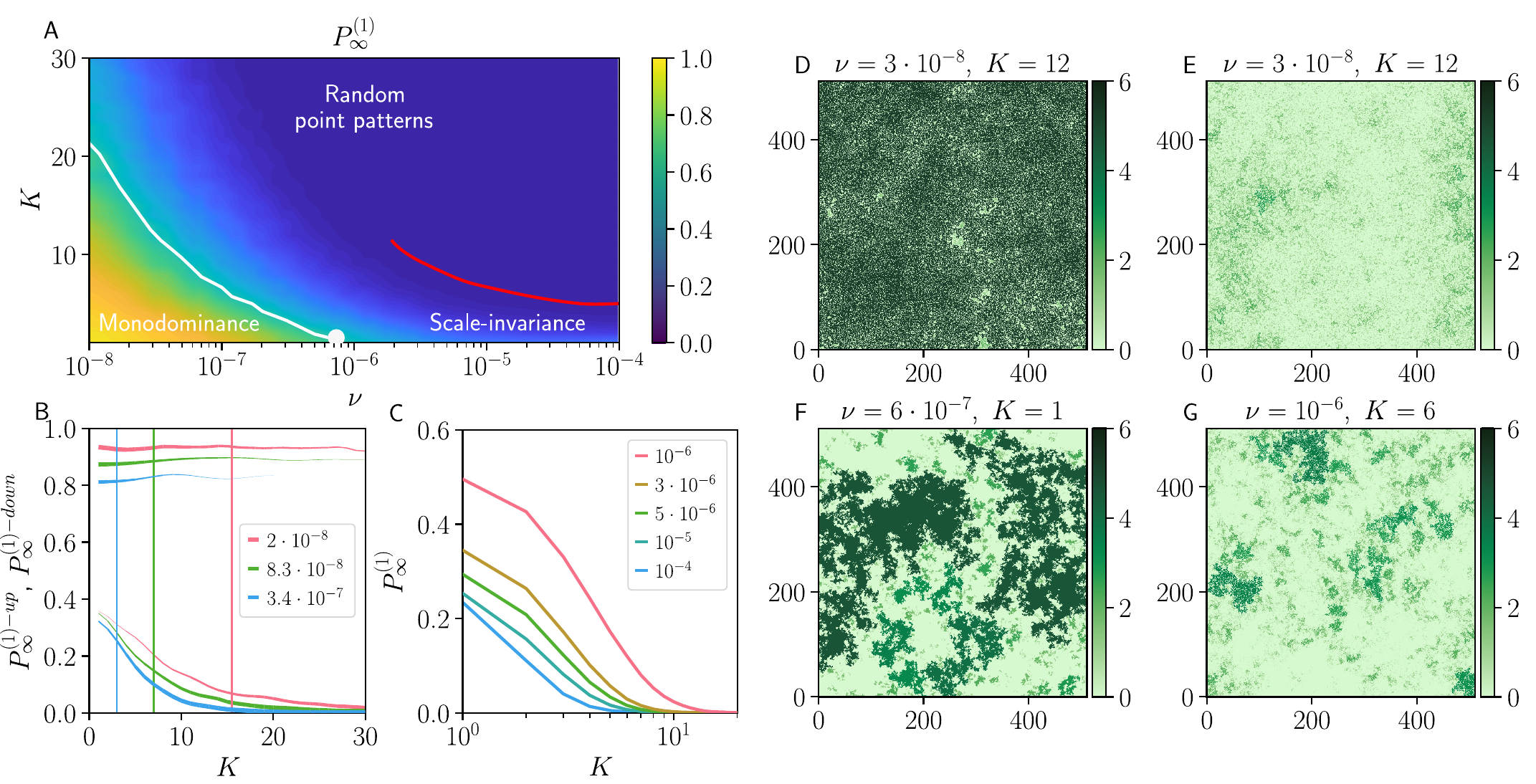}
    \caption{\textbf{(A)} Phase diagram for the most abundant species. Colors represent the intensity of the percolation strength, $P^{(1)}_\infty$, at the single-species level. White line (involving bistability, which vanishes at the white point) divides monodominant states, i.e., where more than 50\% of the tree canopy comprises a single tree species, to coexistence between a small number of species. Redline shows the estimated limit of the scale-invariant regime. Non-homogeneous random point patterns emerge for large dispersal kernels (bluish regions). \textbf{(B)} $P^{(1)}_\infty>0.5$ and $P^{(1)}_\infty<0.5$ vs $K$ for different values of $\nu$ (see legend). The size of each branch is proportional to its relative probability. Vertical solid lines show the Maxwell point for each case. \textbf{(C)} $P^{(1)}_\infty$ vs $K$ for different values of $\nu$ (see legend) in the scale-invariant regime,  characterized by non-vanishing values of $P^{(1)}_\infty$ (see SI2).  \textbf{(D-G)} Clusters of vegetation for the most abundant species accounting for: (D)-(E) Two different realizations in the monodominant regime for high and low values of $P^{(1)}_\infty$, (F) the region when bistability disappears and, (G) the scale-invariant regime.  Color intensity reflects the logarithm of the cluster size. Simulations have been averaged over $10^3$ runs on a lattice of size $N=512^2$.}
     \label{PhaseD}
\end{center}
\end{figure*}

Stochasticity and competition (controlled by local abundances) with other species generate, as in mean-field neutral models \cite{Azaele2016}, a non-equilibrium stationary state where the number of species fluctuates around a mean value, depending on $\nu$. Nevertheless, its mean-field counterpart has no order parameter and no formal phase transition depending on $\nu$ values \cite{Henkel2008}. As mentioned above, the evolution of the population density for the most abundant species can be described as a Langevin equation for a two-species voter model \footnote{From the perspective of a single species, the rest of species are formally equivalent when competing for space and thus, it is possible to consider the dynamics as a two-state model in order to simplify it.} in the presence of a small external driving \cite{Villegas2017,Borile2013}, the mutation rate, that has been proved to remove the process' absorbing states \cite{Borile2013}.

We propose an order parameter capturing essential spatial properties of abundant species: (i) the ability to fill all the available space, i.e., to percolate in space, and (ii) exhibit high conspecific correlations and clustering, i.e., compact clusters. Thus, the collective state can be quantified through the percolation strength concerning the most abundant species, defined as the average size of the largest cluster, $\langle S_M \rangle$, normalized to its total number of trees $N_i$, $P_\infty^{(1)}=\frac{\langle S_M \rangle}{N_i}$ (alternatively, the total system size can be considered, thereby defining $P_\infty$ but without altering the results, see SI1).  Consequently, in an environment characterized by mono-dominance, both values $P_\infty$ and $P_\infty^{(1)}$ are expected to be high, while they can differ in mega-diverse ecosystems, e.g., if an individual species can generate compact clusters without spatial percolation at large scales.


\section*{Results}






 
We have performed a detailed computational study of the SENM \new{--exploiting its duality with coalescing random walks \cite{Bramson1996, Durrett1996, Holley1975}--} depending on both the specific values of $K$ and $\nu$, \new{and averaged over many different and independent realizations}, in order to obtain a highly accurate phase diagram, needed for forthcoming analyses. 
Our results reveal the bistable nature of the dynamical system for low values of $\nu$ (as reflected in Fig.\ref{PhaseD}B, with two different branches for $P^{(1)}_\infty>0.5$ and $P^{(1)}_\infty<0.5$), together with a very rich phase diagram (see Fig.\ref{PhaseD}A). Fig.\ref{PhaseD}A illustrates the existence of different types of emergent spatial point patterns (see also SI1 for different multispecies spatial distributions): Poisson-like non-correlated random patterns (for large dispersal kernels, in the upper part of the diagram, showing an exponentially increasing number of species as $\nu$ grows), monodominance (in the lower left part of the diagram) and a vast region of scale-invariant spatial patterns for local-dispersal kernels (lower-right part). 


In the monodominance regime, the boundary between states (white line in Fig.\ref{PhaseD}a) corresponds to the so-called '\emph{Maxwell point}’, where both the monodominant and spatially diffused states are equally stable. We also computed the probability distribution $\mathcal{P}$ of $P_\infty^{(1)}$, and used it to determine the stationary scalar potential $V_1=-log\mathcal{P}(P_\infty^{(1)})$.  The effective potential exhibits an explicit bistable nature, with a characteristic deep minimum close to the origin --as stochastic systems with multiplicative/demographic noise generally do \cite{MAM1998,NonNormal}-- for low values of $\nu$ and large dispersal distances, $K$ (see SI1). Observe that --where bistability exists-- the maximum is not a singularity just because $\nu>0$ prevents a pure monodominant (absorbing) state from existing. The Maxwell's point displacement is also revealed through extensive computational analyses of the entire system (see Fig.\ref{PhaseD}B): bistability finally vanishes for large enough values of $\nu$ at low dispersal distances (see the white point in Fig.\ref{PhaseD}a and effective potentials in SI1).

\begin{figure}[hbtp]
 \begin{center}
    \includegraphics[width=1.00\columnwidth]{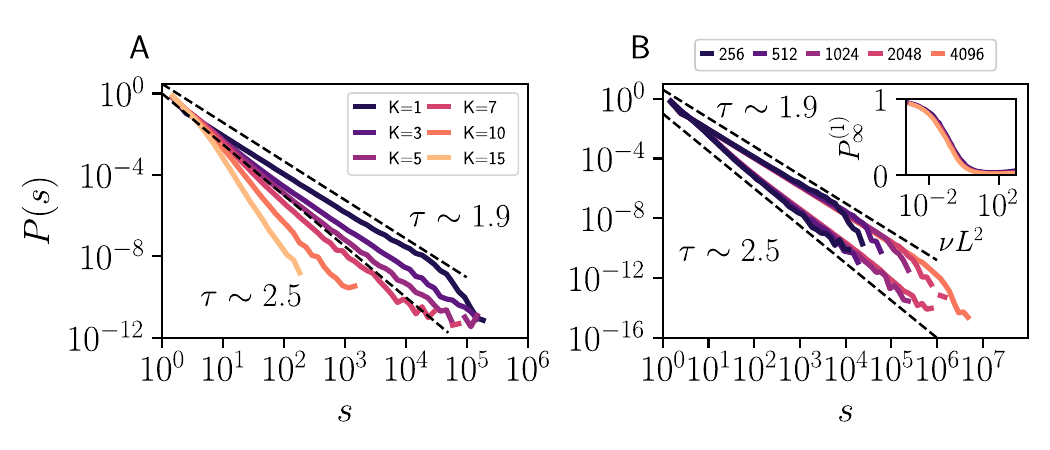}
    \caption{ \textbf{Cluster sizes for the most abundant species.} \textbf{(A)} Cluster-size distributions for $\nu=5\cdot10^{-6}$, $L=512$ for different dispersal kernels. The system exhibits power-law distributed cluster sizes with continuously varying exponents. Black dashed lines are guides to the eye.
    \textbf{(B)}  Finite-size scaling analysis of the cluster size distribution for two kernel values ($K=1$, upper curves and $K=5$, lower curves) and different system sizes (see legend) for a fixed value $\nu L^2=2.62$. Inset: Scaling collapse of $P_\infty^{(1)}$ vs $\nu$ for different system sizes and $K=3$.
    Simulations are averaged over $2\cdot10^3$ realizations.}
     \label{potential}
\end{center}
\end{figure}

Once bistability disappears, non-trivial clusters of vegetation are still present for local-dispersal kernels (see Fig.\ref{PhaseD}F and Fig.\ref{PhaseD}G), characterized by non-vanishing values of $P_\infty^{(1)}$ (see Fig.\ref{PhaseD}C).  To determine the size of contiguous clusters, we describe the distribution of cluster sizes within the von-Neumann neighborhood, i.e., four immediate neighbors without diagonals. Fig.\ref{potential}A and Fig.\ref{potential}B show the distribution of cluster sizes for the scale-invariant regime, showing a scale-free behavior, $P(s)\sim s^{-\tau}$, spanning across many decades and characterized by a variable exponent, which depends on the dispersal kernel (see SI2 for an extensive analysis in terms of $\nu$). More specifically, one can fit exponent values ranging from $\tau=1.9$ to $\tau=2.5$ (fully compatible with percolation critical exponents \cite{Christensen2005,Radicchi2010}). The cluster size distribution becomes exponential for large enough values of the dispersal kernel. Moreover, the system obeys finite-size scaling (as observed at criticality, see Fig.\ref{potential}B) when the immigration probability is properly re-scaled by a factor $\nu L^2$ (see Fig.\ref{potential}C and SI2 for different kernel sizes).

\begin{figure*}[hbtp]
 \begin{center}
    \includegraphics[width=1.8\columnwidth]{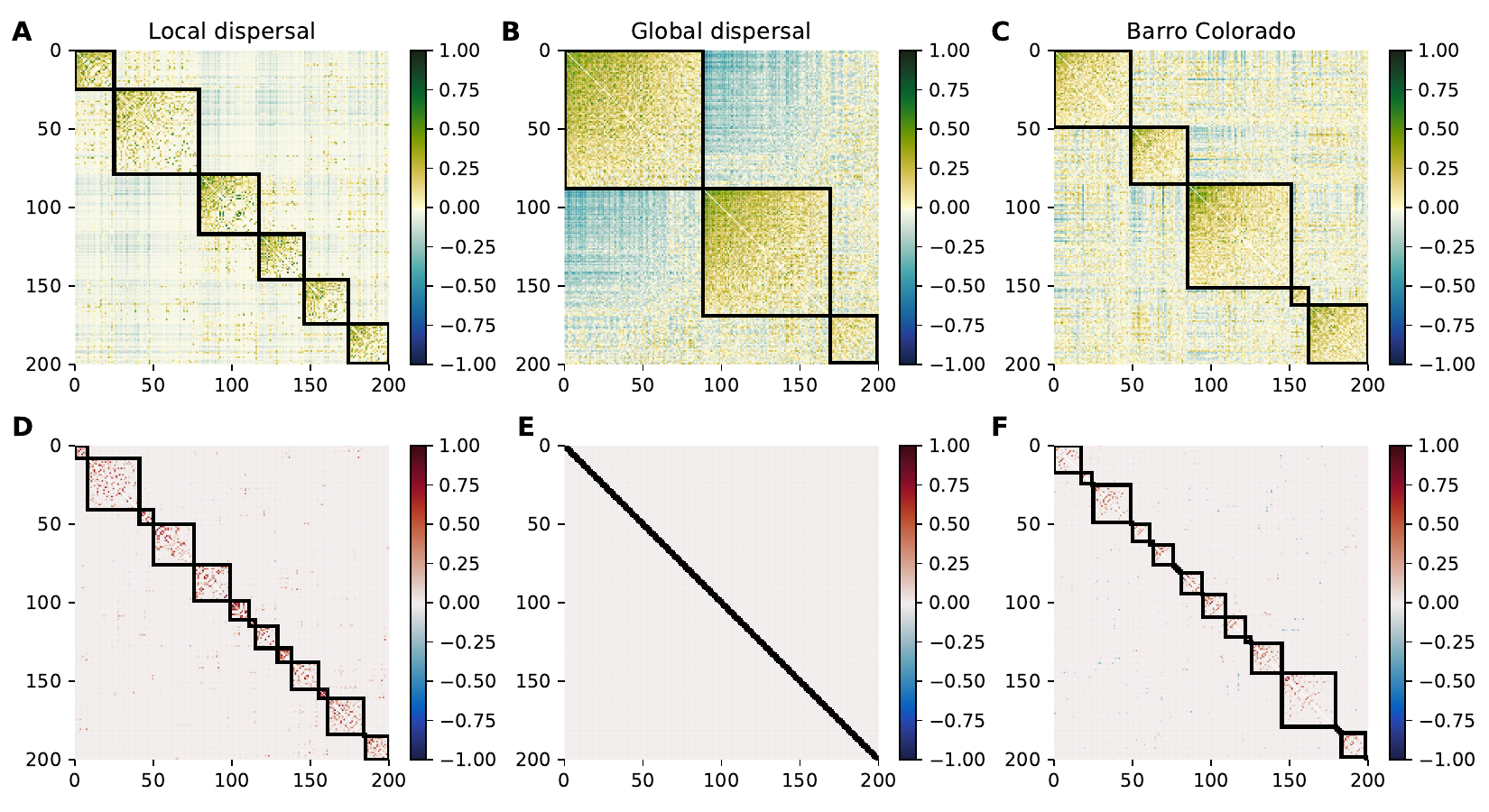}
    \caption{ \textbf{Correlation matrices} Pearson's correlation matrix between species $i$ and $j$ considering the 200 more abundant species in: \textbf{(A)} SENM with local dispersal kernel ($K=5$, $\nu=9.2\cdot10^{-5}$, $L=512$, ${\ell=32}$) \textbf{(B)} SENM with long-distance dispersal ($K=25$, $\nu=3.7\cdot10^{-5}$, $L=512$, ${\ell=32}$) and \textbf{(C)} BCI (${\ell=50m}$). Configurational model for: \textbf{(D)} SENM with local dispersal kernel \textbf{(E)} SENM with long-distance dispersal and \textbf{(F)} BCI. Reorganization in modules is done through the \new{standard} Louvain method over the matrix of positive correlations.  }
     \label{CorrMat}
\end{center}
\end{figure*}

Thus far, we have described the phase space in terms of the most abundant species' properties. However, intricate interactions among species exist in complex environments, and non-trivial interactions are expected to emerge across space \cite{Volkov2009}. These empirical facts motivated us to perform additional analysis of our theory, in which intra-specific correlations are considered.  To do that, we define a protocol to analyze spatial correlations between species. Given N points off different species in an area A, divide the area A in cells, e.g., squares of side $\ell$, and denote with $n^i_\ell(x)$ the number of points for each species in the cell-centered in x. Once this procedure has been done for all species, we computed the averaged Pearson's correlation coefficient between species $i$ and $j$ along with all the boxes. Therefore, the intertwined spatial relationships between species --for a given scale $\ell$-- can be mapped into a correlation matrix between them (as illustrated in Fig.\ref{CorrMat}A-C).

On the one hand, Pearson's correlation matrices between species have been computed for different dispersal kernels. For ease of comparison with empirical results, we selected a specific system size ($N=512^2$) showing a mean value of 300 species (see SI3). Correlation matrices of the top two hundred species exhibit a clear non-trivial modular structure for local dispersal kernels ($K=5$, see Fig.\ref{CorrMat}A), much more diffuse for large dispersal kernels ($K=25$ see Fig.\ref{CorrMat}B). Besides that, correlations between species in BCI --at the macro-scale, i.e., fixing box sizes of $\ell=50m$-- exhibit a non-trivial modular structure between species (see Fig.\ref{CorrMat}C).

On the other hand, null correlation matrices have been generated to assess the modular structure's validity. To do so, we generated null matrices with the same row (or column) sum values as those for the original matrix, using the standard configuration model \cite{Masuda}. Thus, the conservation of the variance --for each node-- of the original correlation matrix remains ensured. We want to point out that this approach is entirely analogous to the configuration model for networks, which preserves the row sum of the adjacency matrix (i.e., degree of each node) \cite{Masuda,Cimini2019,Caldarelli2007}. Figure \ref{CorrMat}D shows that for local dispersal kernels (within the scale-invariant regime), modularity emerges as a robust feature of the system, while for large dispersal kernels (that is, uncorrelated spatial patterns), modularity disappears (see Fig.\ref{CorrMat}E). For the actual case of BCI (Fig.\ref{CorrMat}F), we have also confirmed the modular structure's robustness between species.

Additional analyses of our theory have been done to test the replication ability of diverse empirical facts in Barro Colorado. We computed the pair correlation function, $g(r)$, and the spatial density fluctuations, $\delta_r n$, along with \new{the} associated \new{species abundance distribution (SAD)}. \new{Both quantities have been recently argued to be key to draw conclusions about spatial point patterns \cite{Villegas2020}.} 

\new{The pair correlation function, $g\left(r\right)=\frac{1}{N\rho_{0}2\pi r} \stackrel[i,j]{N}{\sum} \delta\left(r-r_{ij}\right)$, quantifies the average density of trees at distance r from any individual tree, normalized by the mean density of vegetation $(\rho_0)$. For a completely random distribution of points, one expects a flat pair correlation function, $g(r)=1$, by definition. Conversely, values above 1 denote clumping, i.e., tree clustering, which has been generically found in
rainforests \cite{Condit2000}, while values below $1$ indicate anticorrelations. Spatial density fluctuations are based on the same methodology used to build the Pearson's correlation matrix but measure the root mean square deviations, $\delta_rn=[\langle n^2_r(\bm x)\rangle-\langle n_r(\bm x))\rangle^2]^{1/2}$, where $\langle [\dots]\rangle$ indicates the average over all cells. Generically, $\delta_r n\propto \langle n_r\rangle^{\gamma}$, where $\gamma$ ranges in $1/2\leq \gamma \leq 1$. Values of $\gamma>\nicefrac{1}{2}$ are expected to indicate long correlations, a hallmark of out-of-equilibrium systems, and/or the
presence of anomalous spatial heterogeneities \cite{Eisler2008}.  Finally, species-abundance distribution involves all species in a community, ranked from most to least abundant.

\begin{figure}[hbtp]
 \begin{center}
    \includegraphics[width=1.00\columnwidth]{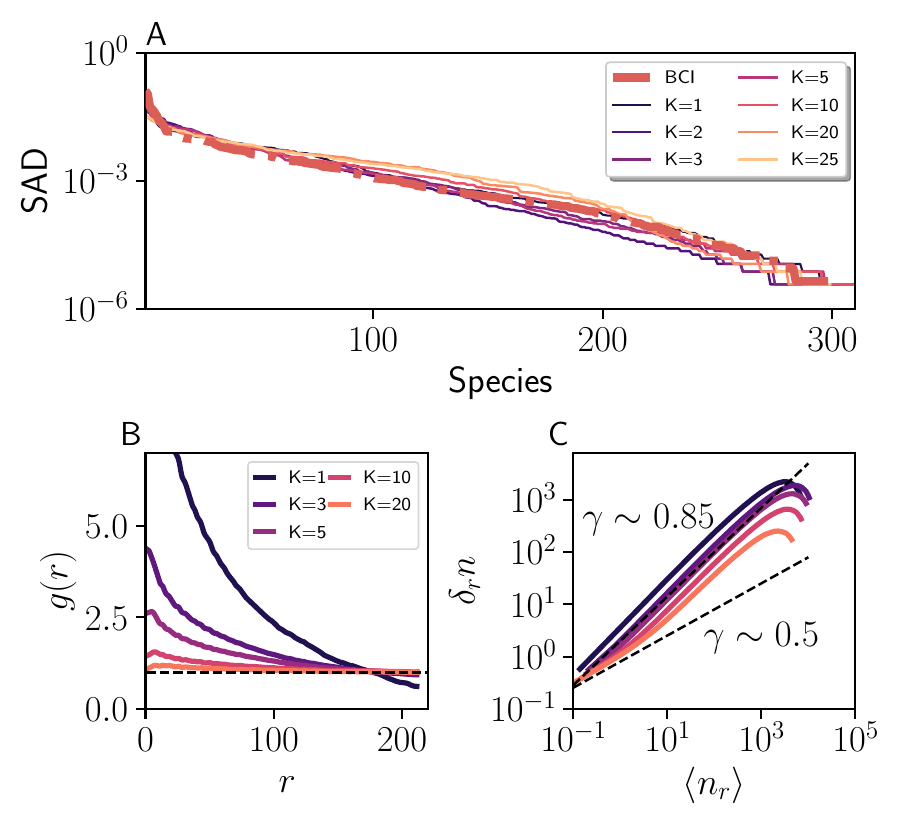}
    \caption{\new{ \textbf{Spatial correlations and fluctuations. (A)} Species abundance curves for different selected realizations of the SENM and different kernel sizes (see legend) along the isoline with $n_s=300$ species, together with the actual rank abundance of the last census of BCI (red dot-dashed line). \textbf{(B)} Pair correlation function, $g(r)$, vs. distance for $\nu=10^{-4}$ and different dispersal kernels, $K$ (see legend) for the most abundant species. Black dashed line corresponds to the result for an utterly homogeneous distribution. \textbf{(C)} Spatial density fluctuations for the most abundant species for $\nu=10^{-4}$ and different dispersal kernels, $K$. Dashed black lines display Taylor's law exponents $\gamma=1/2$ (corresponding to a homogeneous random process), and $\gamma \approx 0.85$ (for the sake of comparison with the observed value in BCI \cite{Villegas2020}). Curves have been averaged over $10^2$ realizations in a system of size $N=512^2$. }}
     \label{AddMeas}
\end{center}
\end{figure}

Fig.\ref{AddMeas}A shows the species abundance distribution for different dispersal kernels in the SENM using the isoline $n_s\simeq300$ species for qualitative comparison with actual data of Barro Colorado Island. There is a good qualitative agreement between the empirical data and the simulated spatial patterns for intermediate values of the dispersal kernel (ranging from $K=3$ to $K=10$). Figure \ref{AddMeas}B and Figure \ref{AddMeas}C  show the spatial correlations and spatial density fluctuations for different values of the dispersal kernel, $K$, in the scale-invariant regime (see SI4 for further analyses depending on $\nu$, and \cite{Villegas2020} for additional explanations). For high values of the dispersal kernel, $K$, and independently of the immigration probability, $\nu$, the pair correlation function is almost one, thus suggesting a random distribution of points in the available space. However, for local dispersal kernels in the scale-invariant regime, $g(r)>1$ values at small distances, together with non-trivial scaling of density fluctuations at all scales, provide evidence of clumping and confirm our previous findings.}
 
 \new{In summary}, high clustering levels at short scales emerge only at the scale-invariant regime, together with anomalous density fluctuations \new{at all scales} and fully compatible species abundance distributions of species \new{regarding current BCI data}. Both quantities' joint assessment justifies the non-trivial emergent correlations among species and suggests the qualitative agreement of BCI data with the scale-invariant regime \cite{Villegas2020}.

\section*{Discussion and conclusions}
Neutral interpretations have revealed to be helpful in the comprehension of a large variety of social and biological scenarios: propagation of memes \cite{Gleeson2014}, microbiome evolution \cite{Zeng2015}, microbial communities \cite{Cira2018}, tumor evolution across cancer types \cite{Williams2016},  causal avalanches in up-states of cortex functioning \cite{Martinello2017}, or the stem cell renewal of the intestinal epithelium \cite{Lopez2010}, to name but a few.

Limited diffusion is expected to play a crucial role in ecological systems. For example, common species are expected to be small-seeded, whereas large-seeded species are consistently rare. However, abundance and seed dispersal do not seem to show a direct relationship \cite{Levine2003}, but there exists evidence of how plants can modify the seed dispersal range to enhance survival \cite{Mcpeek1992}. One could expect spatial effects to be relevant for local diffusion of seeds (far from mean-field behavior, with fluctuations playing an important role \cite{Villa2015,Ricardo2021}), while for large dispersal range, results are expected to be closer to the mean-field case. In particular, a key challenge is to test predictions based on disperser behavior against field data \cite{Nathan2000}.

Despite many efforts on shedding light on how NT can qualitatively explain several patterns observed in ecosystems \cite{Azaele2016}, only some works have contributed to show the ability of its spatially explicit counterpart to reproduce beta-diversity \cite{Chave2002} and species-area relationships \cite{Rosindell2007,Pigolotti2009}.

 Here, we proposed scrutinizing a richer, though a still simple model of spatial neutral competition, including \new{the} dispersal of seeds \cite{Pigolotti2009,Pigolotti2018}. In particular, it exhibits different types of characteristic regimes (i.e., spatial point patterns), as revealed by analyzing the system's order parameter upon changing seed dispersal and immigration parameters.  There is a region of bistability (which exhibits monodominance or uncorrelated point patterns among a small number of species) and a scale-invariant region (shifting to uncorrelated patterns for large dispersal kernels). Meticulous computational analyses allowed us to uncover that, in the scale-invariant regime, scale-free distributed clusters of vegetation emerge, with exponents compatible with percolation transitions \cite{Radicchi2010,Henkel2008,Christensen2005}.


From a global perspective, we have made progress in developing a theoretical framework connecting spatial neutral models (subject to demographic fluctuations), ecological networks, and critical transitions, a \new{key} challenge to understand the crucial interplay between ecological dynamics and species interactions \cite{Azaele2016,Rosindell2011}. In particular, we showed that SENMs are susceptible to reproduce most of the main features observed experimentally in natural rainforests: (i) The \new{species} abundance distribution, (ii) high level of clustering at short scales \cite{Condit2000}, (iii) non-trivial spatial density fluctuations for local dispersal kernels \cite{Villegas2020}, (iv) a modular-like Pearson's correlation matrix between species in the scale-invariant regime, (v) the emergence of scale-free distribution of clusters previously reported in Kalahari \new{woodlands} or tree-canopy gaps in BCI \cite{Sole1995,Scanlon2007} and, (vi) to generate the specific spatial patterns leading to the coexistence mechanism of multiple species also existing in natural rainforests \cite{Wiegand2021}.



Let us emphasize that we have not explored niche effects, which are fundamental to explain  interspecific competitions \cite{Chase2003}. \new{In particular, niche effects as, e.g., competition for diverse limiting resources, response to environmental changes, (co-)evolution of pairs of species (e.g., predator-prey, host-parasite, etc.), among others, can naturally lead to modular interactions between species \cite{Cai2020,olesen2007}}. However, NT omits limiting resources or asymmetric interactions by making the radical assumption of species equivalence (limited dispersal and speciation alone have proven only partially to explain beta-diversity in tropical forests \cite{Condit2002}). The emergent modular structure --\new{despite its complexity}-- reflects only intermittent spatial competition effects (i.e., correlations) \new{changing over time}, and cannot explain bona fide biological interactions among species. Indeed, without seeking to propose any claim of neutrality for actual data, we have chosen this approach just for the sake of simplicity, agreeing with the standard overview: \emph{“niche and neutral models are in reality two ends of a continuum with the truth most likely in the middle”} \cite{Chase2003}.
Quenched and temporal disorder and more sophisticated competition mechanisms (e.g., amensalism and competition for resources) accounting for niche effects, habitat structure \new{(e.g., the emergence of clustering in semi-arid ecosystems \cite{Scanlon2007})}, and species differences will be analyzed in future work.

However, even if the SENM studied here is exceedingly simple to be a realistic model of a complex ecological landscape, it can provide us with insight into the basic dynamical mechanisms needed to generate its complex dynamical features, paving the way to the long term goal of constructing a statistical-mechanics of rainforests. Furthermore, we propose a novel, powerful and easy-to-use method for building networks in multi-agent dynamic systems. We also believe that our approach can open the door to novel research lines in the context of bacterial communities or the spreading of opinion dynamics.



%
%


\begin{acknowledgments}
 GC acknowledges Israel-Italy collaboration project Mac2Mic and EU Grant nr. 952026 "Humane-AI-Net" for financial support. PV thank M. Cencini, M.A. Muñoz, L.Falsi and V.Buendia for very valuable discussions and useful comments.
 \end{acknowledgments}

\vspace{-0.5cm}

\def\url#1{}
\def\bibpreamble{\vspace{-0.5cm}*Corresponding author. \\ pvillegas@ugr.es \\}
%

\clearpage


\section*{Supplementary material}

# Supplementary Information: Emergent spatial patterns of coexistence in species-rich plant communities

Pablo Villegas*,[1] Tommaso Gili,[1] and Guido Caldarelli[2, 3, 4, 5]

[1]*IMT Institute for Advanced Studies, Piazza San Ponziano 6, 55100 Lucca, Italy.*
[2]*Department of Molecular Sciences and Nanosystems,*
*Ca' Foscari University of Venice, 30172 Venice, Italy*
[3]*European Centre for Living Technology, 30124 Venice, Italy*
[4]*Institute for Complex Systems, Consiglio Nazionale delle Ricerche, UoS Sapienza, 00185 Rome, Italy*
[5]*London Institute for Mathematical Sciences, W1K2XF London, United Kingdom*

**CONTENTS**

# 1. THE SPATIALLY EXPLICIT NEUTRAL MODEL

## 1.1. Nearest Neighbors

The 2D-SENM (widely known as the multispecies voter model) is usually simulated using a nearest-neighbor dispersal kernel. The dispersal kernel is thus automatically fixed, and the dynamics of the system depend only on one parameter: the immigration probability $\nu$. As previously demonstrated in [6], in the mean-field case, the system exhibits a bifurcation at some value of $\nu_c$ –which is not a proper thermodynamic phase transition– from a unimodal distribution (coexistence) to a bimodal one. As shown in Fig. 1 for the two-dimensional system, where different spatial distributions for different species are shown, it can be better understood in the following terms: at low values of $\nu$, the monodominant state is the most stable one (see, e.g., Fig. 1A) but, for a precise value of $\nu_c$, the most abundant species lose the ability to generate an infinite cluster invading the entire space, making thus possible the stable coexistence of multiple species (see Fig. 1C and Fig. 1D).

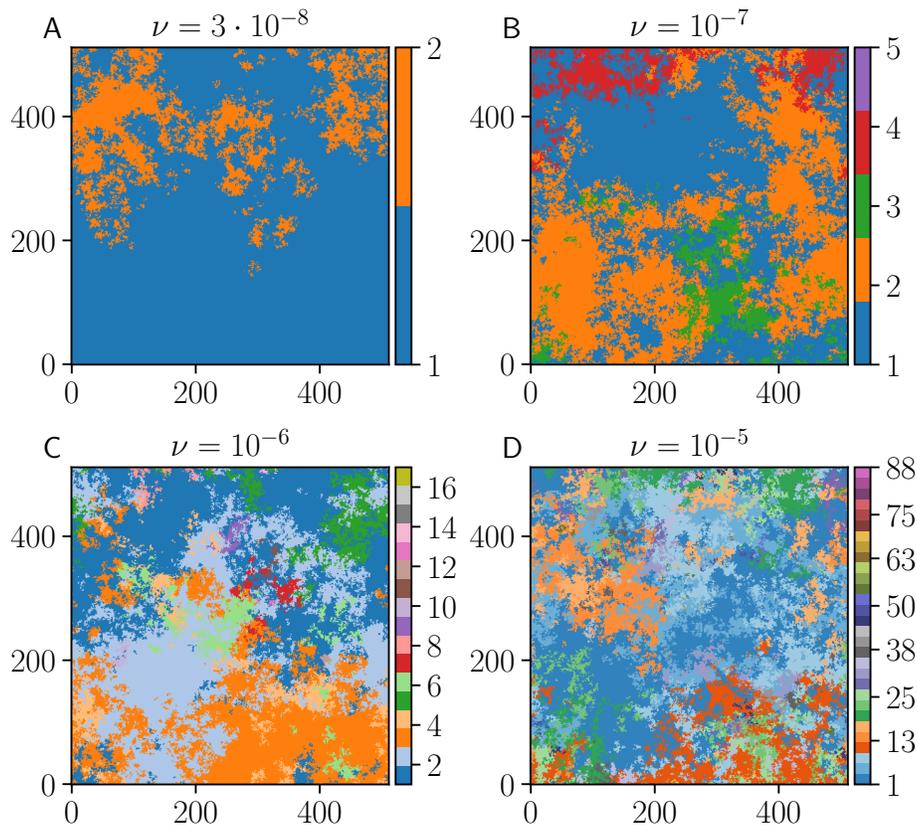

FIG. 1: **Spatial patterns for the SENM with NN kernel.** Selected realizations of the SENM for different values of $\nu$ (see figures title): (A)-(B) In the monodominant phase (C) a close-to $\nu_c$ value (observe that the most abundant species cannot generate a giant cluster in this case) and (D) multispecies coexistence. Each different color stands for a different species, ordered by abundance. The dark blue color indicates the most abundant species for all cases. Near some specific value of $\nu$, new species' growth makes impossible to the most abundant species to generate an infinite cluster invading all space. Beyond this point, high-complex spatial patterns coexist in space. Observe that clusters exhibit a compact structure for all cases. Simulations have been performed for a squared lattice of size $L = 512^2$

## 1.2. Dispersal Kernel

We present here computational analyses of the infinite cluster in the whole space (i.e., normalizing the total size of the infinite cluster for the most abundant species by the total number of lattice points, instead of its total number of trees), $P\infty$. The phase diagram is in complete agreement with the individual species one. As shown in Fig.2 the computationally-obtained phase diagram exhibits a qualitatively identical structure to the one-species phase diagram shown in the main text. In a nutshell, the three main phases prevail: a monodominant regime (for which the most abundant species usually occupy more than half of the total space), a scale-invariant one, and a random regime characterized by global diffusion without long-correlations between individuals.

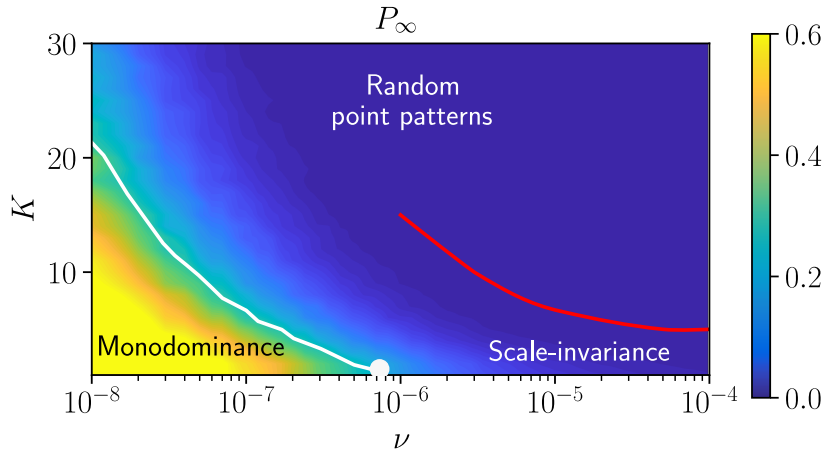

FIG. 2: **Phase diagram for a 2D lattice of interacting species** for different values of $K$ and $\nu$, using the global infinite cluster $P_\infty$ as order parameter (color-coded). Yellowish colors stand for monodominant states, while blueish colors correspond to non-percolating states. The monodominance regime clearly shows that the most abundant species exhibits an infinite compact cluster in space. As in the main text, the white line represents the bistability line, the white point describes the coordinates for which bistability disappears, and the red line shows the estimated limit of the scale-invariant region.

Even though the system's main phases have been carefully presented in previous sections, we still do not have a transparent explanation for the system's observed states nor a detailed analysis of how the phenomenon depends on dispersal kernel. Aiming to shed further light on these issues, we present a compelling description of the entire system dynamics in this supplementary section.

To do that, we have computed the probability distribution of the infinite cluster, $P_\infty^{(1)}$, at the single species level, defining an effective 1D potential $V(P_\infty^{(1)}) = -log(P(P_\infty^{(1)}))$. Fig.3 illustrates the potential for different values of $\nu$ and different values of the kernel size, $K$. Observe, in particular, that the effective potential displays a bistable nature for low values of $\nu$, with a well-defined minimum around the monodominant case. In contrast, for bigger kernel sizes, a new minimum emerges for non-clustered dispersion of the species. For local dispersal kernels, the higher the values of $\nu$, the smaller the monodominant state's relative stability (see Fig.3), until it completely disappears –corresponding with the white point in Fig.2– giving rise to a non-zero value of $P_\infty^{(1)}$. For high values of $\nu$ and high values of $K$, the potential exhibits a single-peaked shape with its minimum centered at small values of $P_\infty^{(1)}$, a signature of non-correlated spatial patterns driven by diffusion.

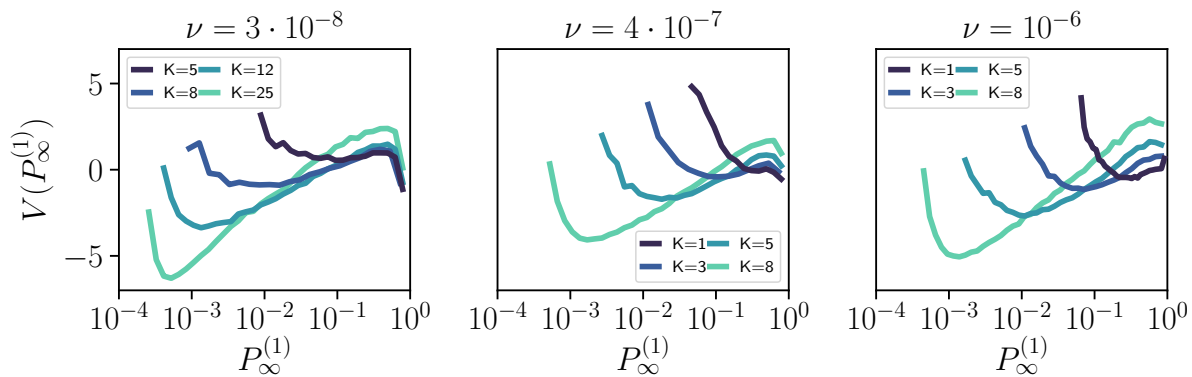

FIG. 3: **Effective potentials** for the infinite cluster, $P_\infty^{(1)}$, and different values of the kernel size ($K$, see legend) and different values of the immigration probability, $\nu$ (see title). The system exhibits bistability for low enough values of $\nu$, as revealed the double-well nature of the potential. After some critical value of $\nu_c$, a single-peaked potential appears, with a well-defined minimum placed into shrinking values of $P_\infty^{(1)}$ when $K$ increases. Effective potentials have been computed averaging over $10^4$ independent realizations.

A crystal-clear picture of the system's different phases can be shown by illustrating different snapshots at different phase diagram points. As can be seen in Fig.4A and Fig.4B, for low values of $\nu$, the predominant species always fills all the available space with clusters becoming more and more diffused for growing values of the dispersal kernel (see also Fig.5 and Fig.6). On the contrary, high values of $\nu$ promote species diversity, as reflected in the high number of coexisting species in Fig.4D, Fig.5D, and Fig.6D. However, in this particular case, high conspecific correlations are ensured only for local dispersal kernels (see the particular case of $K = 5$ in Fig.4D), obtaining random heterogeneous patterns for large dispersal kernels (see below). Interestingly, the intermediate region (see Fig.4C and Fig.7) involves a relatively small number of species ($\sim$ 10-20), where both maximum spatial correlations and conspecific correlations are maximized, without necessarily having compact clusters for the individual species. This specific region, which needs to be carefully scrutinized in the future, is no more than the transition between the bistable regime to the scale-invariant one.

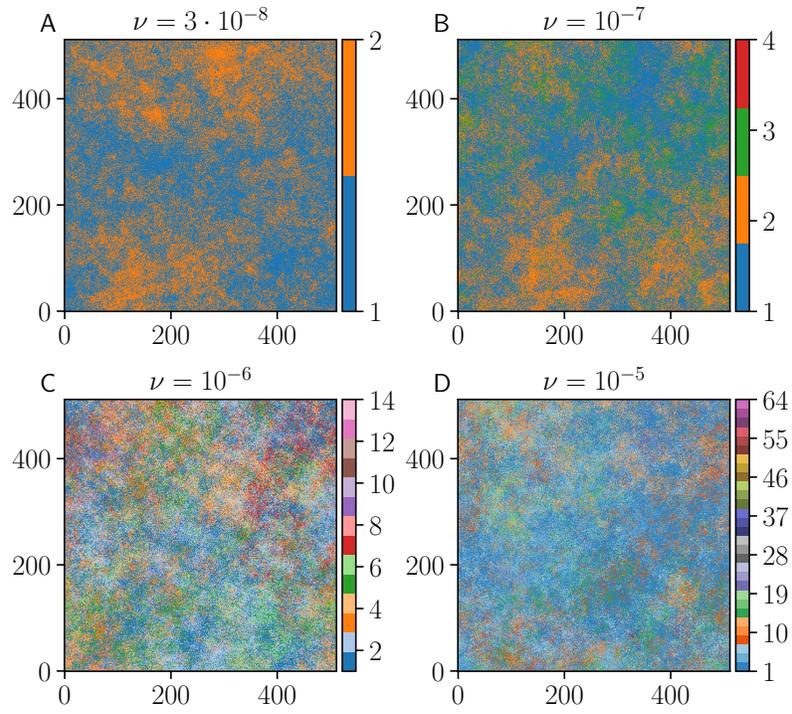

FIG. 4: **Snapshots for K=5 and growing values of** $\nu$

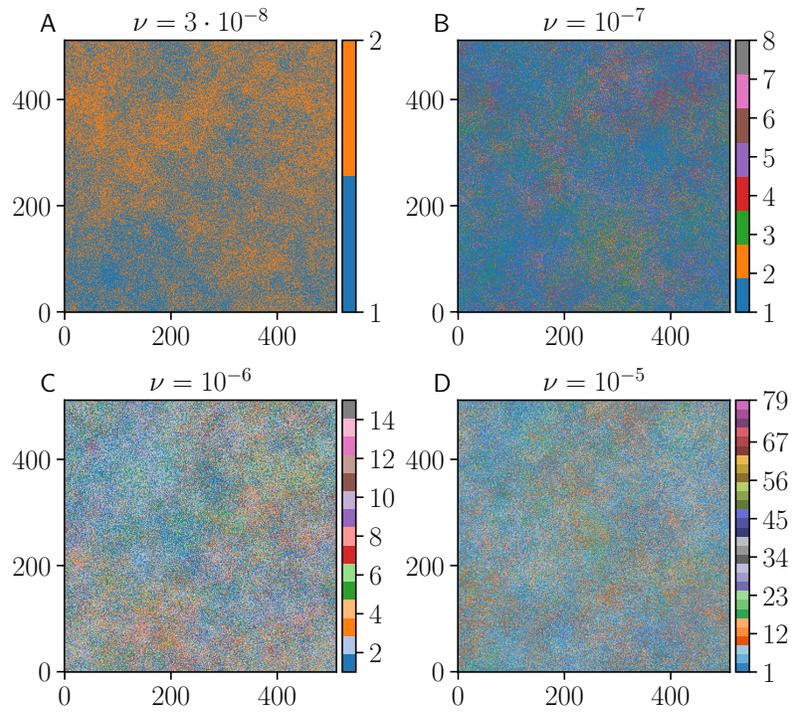

FIG. 5: **Snapshots for K=10 and growing values of** $\nu$

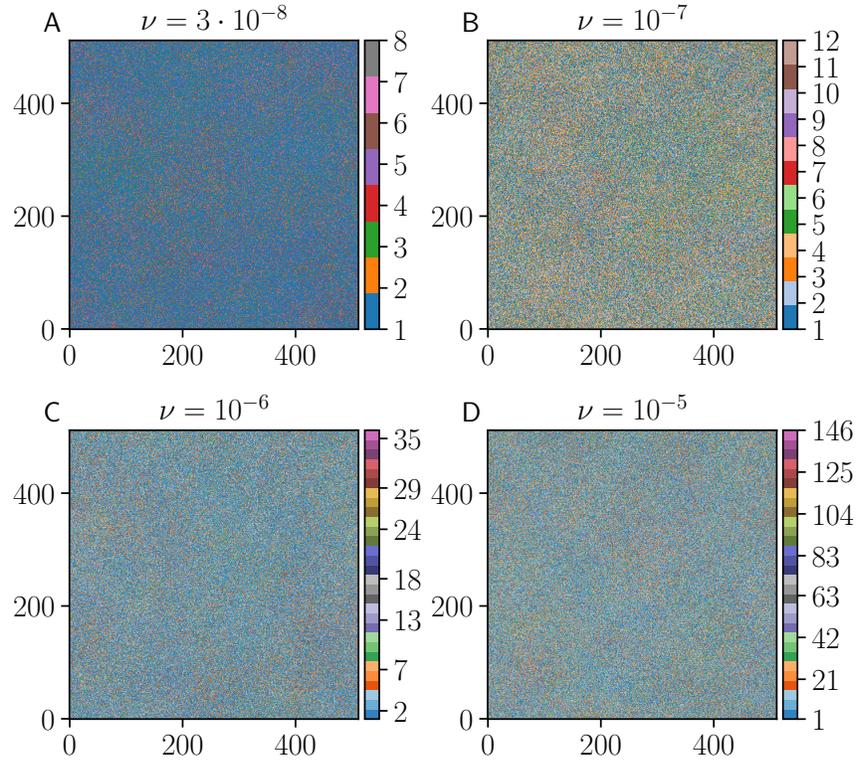

FIG. 6: **Snapshots for K=25 and growing values of** $\nu$

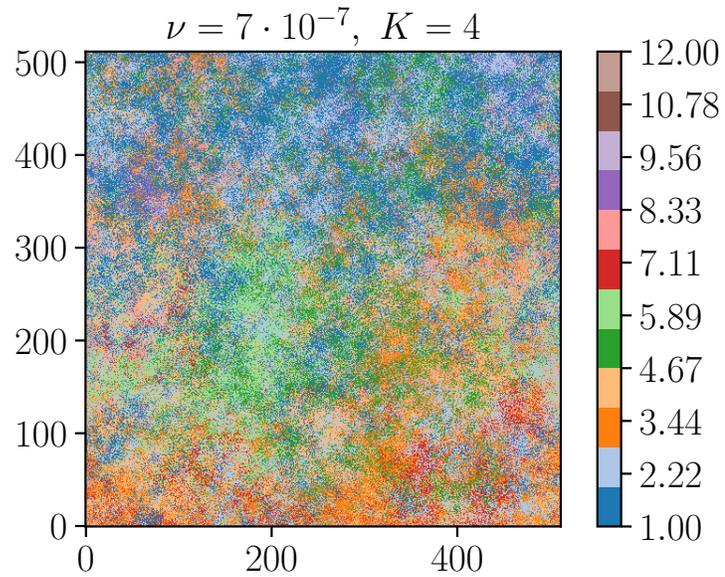

FIG. 7: **Snapshot** in the region where bistability disappears

## 2. SCALE-INVARIANCE ON THE SENM

Here, we further investigate whether scale-free clusters of vegetation for the most abundant species emerged when we moved along the scale-invariant regime. For a specific value of $\nu$, results for different dispersal kernels are reported in Fig.2 of the main text, displaying scale-free distributed clusters only for local dispersal kernels. Here, we show results for different values of $\nu$ slightly below and above this value. As discussed at the extent in the main text, Fig.8 shows that scale-free clusters of vegetation (with power-law distributed sizes) emerge in all the scale-invariant regime with variable exponents. In particular, one can see that the increase of the total number of species (i.e., considering high values of $\nu$), with further spatial competition, lead to a limited spatial diffusion at the individual species level, thus constraining the total number of individuals, but not the scale-free distribution of vegetation clusters.

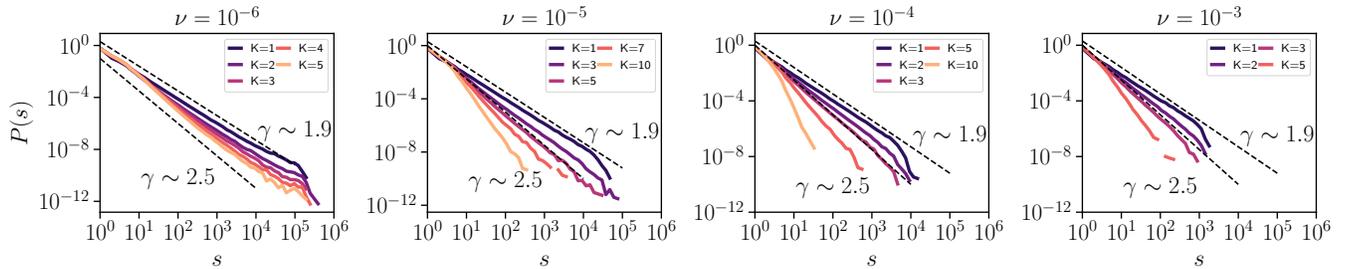

FIG. 8: **Cluster size distribution** for different values of $\nu$ (see title) and different dispersal kernels (see legend). The most abundant species exhibits power-law distributed clusters with variable exponent for all the scale-invariant regime, ranging from 1.9(1) to 2.5(1). Conversely, as expected for large dispersal kernels, the random regime is characterized by an exponential cluster sizes distribution with a well-defined scale.

Moreover, a careful inspection of the immigration probability reveals a clean finite-size scaling for different dispersal kernels so that the cutoffs of cluster size distributions increase in a scale-invariant way upon enlarging the system size (see main text). Another unequivocal evidence of such scaling is the clean collapse of the order parameter, $P_\infty^{(1)}$, shown in Fig.9. Observe that, for diverse values of $K$, all curves collapse under the consideration of the rescaled immigration parameter $\tilde{\nu} = \nu L^2$, but showing a system size dependence in the scale-invariant regime. It can, therefore, be said that the phase diagram presented in the main text is scale-independent when the one-place immigration probability is considered.

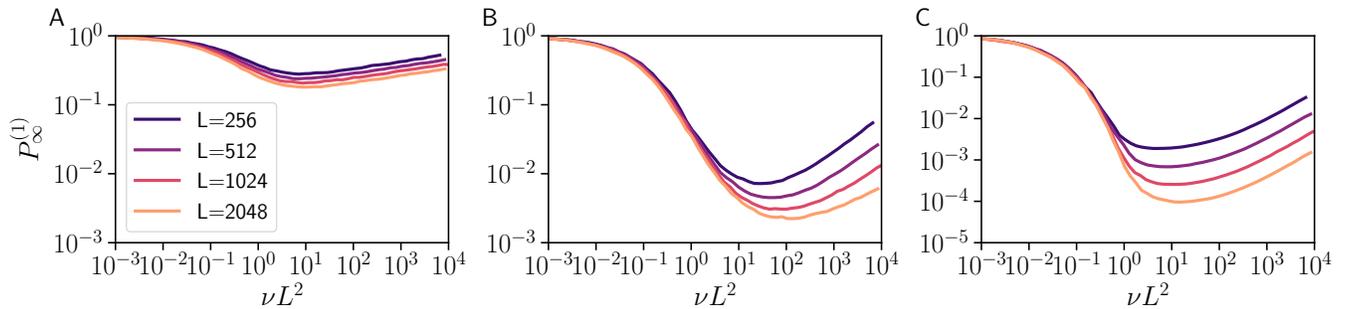

FIG. 9: **Scaling collapse for different system sizes** (see legend) and different dispersal kernels, namely: (A) $K = 1$, (B) $K = 5$ and (C) $K = 10$.

## 3. NUMBER OF COEXISTING SPECIES

For given values of $\nu$ and $K$ the number of species in the ecosystem exhibit a statistically steady value [4, 5] when averaged over different realizations (i.e., fluctuating around a single-peaked parabolic potential with a well-defined mean value). However, notice that, because of speciation, the total number of species is not fixed a priori. As a complementary measure, we have computed the scaling of the total number of species for different system sizes once the immigration probability by lattice point is fixed, i.e., by considering $\tilde{\nu} = \nu L^2$ for different kernel sizes. Fig.10 shows the mean number of species as a function of the system size for different values of $\tilde{\nu}$, showing that it remains constant for low values of $\nu$, starting to grow when bistability vanishes slightly.

An extensive analysis of the mean number of species as a function of $K$ and $\nu$ (both in linear and semi-log scale) is shown in Fig.11 for a system size $N = 512^2$. As expected, the total number of species grows fast as $\nu$ increases, while for growing dispersal kernel, a moderate increase in the total number of coexisting species can be observed.

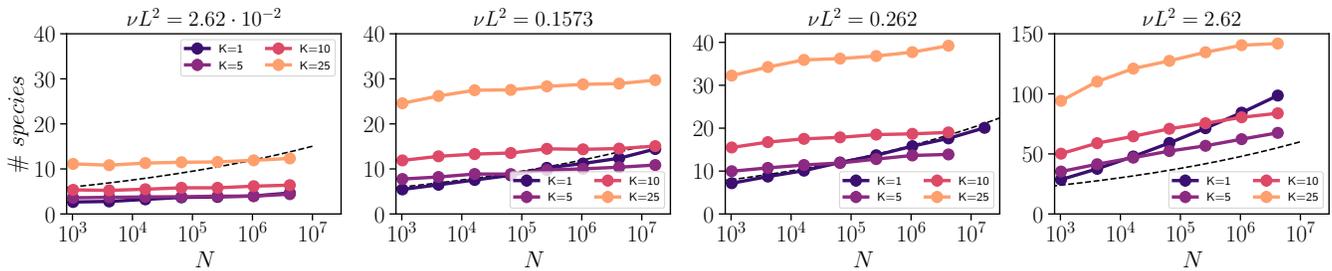

FIG. 10: **Scaling of the total number of species.** Mean number of species versus system size for different values of $\nu$ (see title) and different kernel sizes (see legend). The black dashed line, which serves as a guide to the eye, represents a growth with the functional form $N^{0.1}$. Also, outside of the monodominant regime (leftmost figure), the case with a low dispersal kernel, $K = 1$, exhibits an entirely different behavior to all other cases. Error bars are smaller than the symbols.

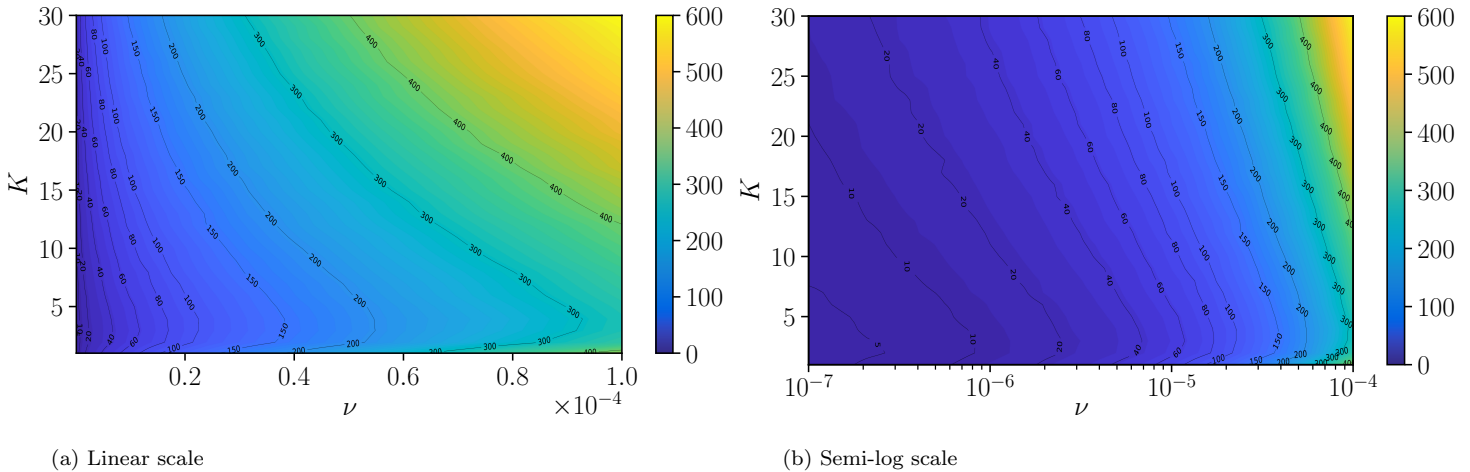

(a) Linear scale

(b) Semi-log scale

FIG. 11: Mean number of coexisting species in the SENM for a system size $N = 512^2$. Color intensity represents the total number of coexisting species (see color bar). Black solid lines are isolines with a fixed mean number of species.

## 4. ANALYSIS OF SPATIAL CORRELATIONS AND DENSITY FLUCTUATIONS

Recently, it has been argued the need to consider both the pair correlation function, $g(r)$, and the spatial density fluctuations, $\delta_r n$ to draw conclusions about spatial point patterns [7]. We perform an extensive analysis of both quantities, which allows us to state better the nature of the diverse phases of the SENM.

The pair correlation function, $g(r)$, quantifies the average density of trees at distance r from any individual tree, normalized by the mean density of vegetation. Given $N$ points in an area $A$, $g(r)$ corresponds to the contained number of points $dN(r)$ in the annular area between $r$ and $r + dr$ centered on one point, and averaged over all points. Finally, it is normalized with the expected number of neighbours for a completely random (homogeneous Poisson) distribution, i.e., the mean density $\rho_0 = N/A$. Thus, the $g(r)$ reads

$$g\left(r\right) = \frac{dN(r)}{2\pi\rho_0 r dr} = \frac{1}{N\rho_0 2\pi r} \sum_{i,j}^{N} \delta\left(r - r_{ij}\right) ,\qquad(1)$$

where $i$ and $j$ denote two points in the area of study. For a completely random distribution of points, one expects a flat pair correlation function, $g(r) = 1$, by definition. Conversely, values above 1 denote clumping, i.e., tree clustering, which has been generically found in rainforests [8], while values below 1 are indicative of anticorrelations.

Figure 12 shows the pair correlation function, $g(r)$, for different values of $\nu$ (see title) and different values of the dispersal kernel, $K$. For high values of the dispersal kernel, $K$, and independently of the immigration probability, $\nu$, correlations are almost one, thus suggesting a random distribution of points in the available space. Clumping emerges, however, in a clear way, in the scale-invariant regime for local dispersal kernels (see Fig.12).

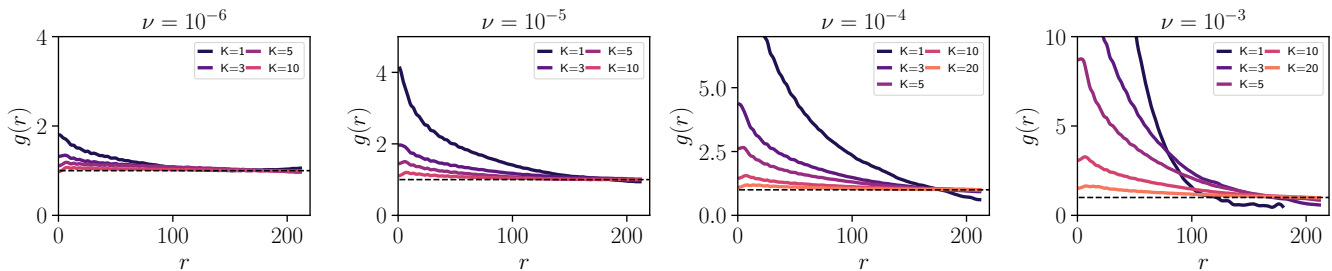

FIG. 12: Pair correlation function, $g(r)$, vs. distance for different values of the immigration probability $\nu$ (see title) and different dispersal kernels, $K$ (see legend) for the most abundant species. $g(r)$ suggests the emergence of random non-correlated patterns both for small values of $\nu$ and high values of the dispersal kernel, $K$. Tree clumping (i.e., values of $g(r) > 1$ at short scales) emerges as a robust feature of the SENM at the scale-invariant regime. Black dashed line corresponds to the result for an utterly homogeneous distribution. Curves have been averaged over $10^2$ realizations in a system of size $N = 512^2$.

Besides that, spatial density fluctuations are measured in the following way: given $N$ points in an area $A$, divide the area $A$ in cells, e.g., circles or squares of side $r$ (properly taking into account the borders [7]), and denote with $n_r(\boldsymbol{x})$ the number of points in the cell-centered in $\boldsymbol{x}$, computing its root mean square deviations, $\delta_r n = [\langle n_r^2(\boldsymbol{x}) - \langle n_r(\boldsymbol{x})\rangle^2]^{1/2}$, where $\langle \dots \rangle$ indicates the average over all cells. For a completely random (homogeneous Poisson) process, $\delta_r n$ is expected to scale as the square root of the mean number of points, i.e.,

$$\delta_r n \propto \langle n_r\rangle^{1/2}\qquad(2)$$

Conversely, it is an empirical observation that for many ecological processes

$$\delta_r n \propto \langle n_r\rangle^{\gamma}\qquad(3)$$

with an exponent $\gamma$ typically ranging in $1/2 \leq \gamma \leq 1$. The power-law behavior (3) is known in the literature as Taylor's Law (TL), and it was first put forward in the context of population ecology [9] (see Ref. [10] for a review). Values of $\gamma > 1/2$ are expected to be indicative of long correlations, a hallmark of out-of-equilibrium systems, and/or the presence of anomalous spatial heterogeneities [10].

Figure 13 shows the scaling of the most abundant species for different values of $\nu$ and different kernel sizes $K$. Note that almost all the possible kernel sizes and immigration probabilities exhibit an exponent indicating deviations from

the homogeneous Poisson process. However, in the specific case of high dispersal kernels, even if anomalous exponents can be observed (with an exponent compatible with $\gamma = 1$), it can be associated with trivial heterogeneities due to the presence of density gradients originated by the diffusion process (once fluctuations are evaluated together with the $g(r)$, which is completely flat). Only at the scale-invariant regime (i.e., intermediate values of $\nu$ and local dispersal kernels) anomalous density fluctuations emerge jointly with a non-trivial pair correlation function. Furthermore, anomalous density fluctuations show a compatible exponent with the observed in actual data, $\gamma \simeq 0.85(5)$ [7] for local dispersal kernels.

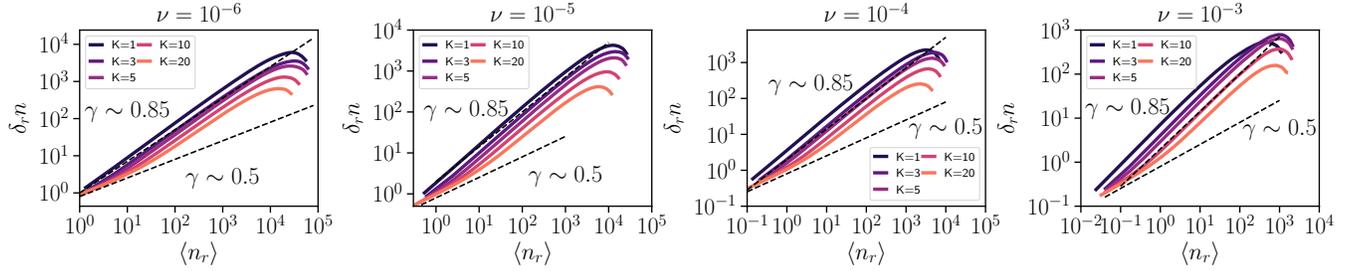

FIG. 13: Spatial density fluctuations for the most abundant species: standard deviation of number of trees, $\delta_r n$, as a function of the mean number of trees, $\langle n_r \rangle$. Dashed black lines display Taylor's law exponents $\gamma = 1/2$ (corresponding to a homogeneous random process), and $\gamma \approx 0.85$ (as a guide to the eye, for the sake of comparison with the observed value in BCI [7]). Curves have been averaged over $10^2$ realizations in a system of size $N = 512^2$.



\end{document}